\documentclass[twocolumn,floatfix,showpacs]{revtex4}
\usepackage{graphicx}
\usepackage{amsmath}
\usepackage{amssymb}
\usepackage{bm}

\begin{document}

\title{Hartman effect and spin precession in graphene}
\author{R. A. Sepkhanov}
\affiliation{Instituut-Lorentz, Universiteit Leiden, P.O. Box 9506, 2300 RA Leiden, The Netherlands}
\author{M. V. Medvedyeva}
\affiliation{Instituut-Lorentz, Universiteit Leiden, P.O. Box 9506, 2300 RA Leiden, The Netherlands}
\author{C. W. J. Beenakker}
\affiliation{Instituut-Lorentz, Universiteit Leiden, P.O. Box 9506, 2300 RA Leiden, The Netherlands}
\date{October 2009}
\begin{abstract}
Spin precession has been used to measure the transmission time $\tau$ over a distance $L$ in a graphene sheet. Since conduction electrons in graphene have an energy-independent velocity $v$, one would expect $\tau\geq L/v$. Here we calculate that $\tau<L/v$ at the Dirac point (= charge neutrality point) in a clean graphene sheet, and we interpret this result as a manifestation of the Hartman effect (apparent superluminality) known from optics.
\end{abstract}
\maketitle

\section{Introduction}
\label{intro}

The precession of the electron spin in a magnetic field provides a clock for the study of the electron dynamics \cite{But02}. This so-called Larmor clock \cite{Baz67,Ryb67} is a particularly useful tool in a quasi-two-dimensional system, when one can use a parallel magnetic field to avoid perturbing the dynamics by the Lorentz force. The single-atomic layer of carbon atoms known as graphene is the ultimate two-dimensional system \cite{Gei07}. Spin precession was used successfully by Van Wees and collaborators to measure the diffusion time through a disordered graphene sheet \cite{Tom07,Pop09}. For a mean free path $l$ small compared to the separation $L$ of the source and detector contacts,  the diffusion time $\tau_{D}\simeq \tau_{0}L/l$ is larger by a factor $L/l$ than the ballistic time of flight $\tau_{0}\simeq L/v$, with $v=10^{6}\,{\rm m/s}$ the energy independent Fermi velocity in graphene.

\begin{figure}[tb]
\centerline{\includegraphics[width=0.9\linewidth]{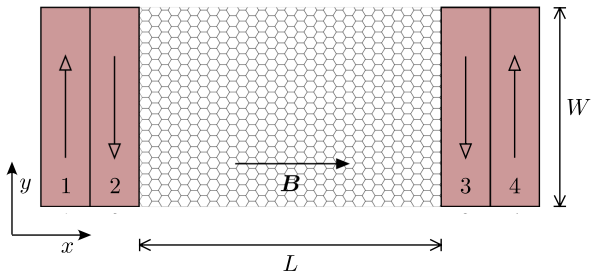}}
\caption{\label{fig_setup}
Schematic top view of a graphene sheet with four ferromagnetic contacts numbered $1,2,3,4$; arrows indicate the direction of magnetization. The ratio $I_{31}/I_{41}$ of currents from contact $1$ into contacts $3$ and $4$ measures the spin precession time $\tau^{(2)}$ in an in-plane magnetic field $\bm{B}$. An alternative geometry, with the magnetization in contacts $3,4$ aligned perpendicularly to the magnetization in contacts $1,2$ (and still perpendicularly to $\bm{B}$), measures the time $\tau^{(1)}$ through the ratio $(I_{41}-I_{31})/(I_{41}+I_{31})$.  
}
\end{figure}

In a clean graphene sheet, when $l\gg L$, the diffusive dynamics becomes ballistic, at least for Fermi energies $\varepsilon_{F}$ away from the Dirac point ($\varepsilon_{F}=0$). At the Dirac point the dynamics in graphene is called ``pseudodiffusive'': conductivity and shot noise suggest diffusive transport even in the absence of any disorder \cite{Two06}. In this paper we theoretically address the question of what spin precession can tell us about the dynamics at the Dirac point.

While the notion of pseudodiffusive dynamics might suggest a scaling $\tau\propto L^{2}$ for the transmission time $\tau$ at the Dirac point, such quadratic scaling is forbidden by dimensional arguments. In the absence of disorder there is only a single length scale $L$ at $\varepsilon_{F}=0$, so $\tau={\rm constant}\times L/v$ is the only quantity with dimensions of time. As we will show, the proportionality constant is $<1$, so $\tau<L/v$ --- as if electrons could propagate at speeds $>v$.

The optical analogue of this anomalously short transmission time, with $v$ replaced by the speed of light, is called superluminality or the Hartman effect \cite{Har62,note1}. As explained by Winful \cite{Win06}, there is no violation of relativity because the transmitted waves are not propagating but evanescent. Graphene would offer an interesting possibility to observe this paradoxical effect in the solid state.

In the next sections we formulate the scattering problem in a clean graphene sheet at the Dirac point \cite{Two06,Kat06}, and calculate the transmission time $\tau$ measured in a weak-field spin precession experiment over a distance $L$. We then perform a separate calculation of the mode-dependent Wigner-Smith delay time $\tau_{n}$, which is directly defined in terms of the scattering matrix \cite{Wig55,Smi60} (without reference to spin precession). This is the quantity studied in the optical context. 

We demonstrate that $\tau$ is the weighted average of $\tau_{n}$, weighted with the mode-dependent transmission propability $T_{n}$. More precisely, depending on the relative alignment of the magnetization at the two ends of the graphene sheet, the precession experiment measures either $\tau^{(1)}$ or $\tau^{(2)}$, defined by
\begin{equation}
\tau^{(p)}=\left[\frac{\sum_{n}\tau_{n}^{p}T_{n}}{\sum_{n}T_{n}}\right]^{1/p},\;\;p=1,2.\label{tautau_nrelation}
\end{equation}
For a graphene sheet with a large aspect ratio (width $W$ $\gg$ length $L$) we calculate
\begin{equation}
\tau^{(1)}=\frac{7\zeta(3)}{\pi^{2}}\frac{L}{v}=0.85\,\frac{L}{v},\;\;\tau^{(2)}=0.87\,\frac{L}{v}.\label{tauresult}
\end{equation}
Both times are below $L/v$, as a manifestation of the Hartman effect.

\section{Spin precession through a graphene sheet}
\label{spinprecession}

We study the four-terminal geometry \cite{note2} of Fig.\ \ref{fig_setup}, in which spin-up electrons are injected into a graphene sheet from ferromagnetic contact $1$ at an elevated voltage $V_{1}$, and drained to ground via three other ferromagnetic contacts $2,3,4$. The two contacts at the same side of the graphene sheet have antiparallel magnetizations. In the existing experiments \cite{Tom07,Pop09}, the contacts at opposite sides of the graphene sheet are collinear. This is the geometry shown in Fig.\ \ref{fig_setup}, where the magnetizations in all four contacts are aligned along the $\pm y$-direction. We will consider this case first, and show that it measures the time $\tau^{(2)}$ of Eq.\ \eqref{tautau_nrelation}. 

The time $\tau^{(1)}$ is measured if the magnetizations in contacts $3$ and $4$ are aligned perpendicularly to those in contacts $1$ and $2$ (along the $\pm z$-direction of Fig.\ \ref{fig_setup}). We defer a discussion of that geometry to Sec.\ \ref{sec_tau1}.

\begin{figure}[tb]
\centerline{\includegraphics[width=0.9\linewidth]{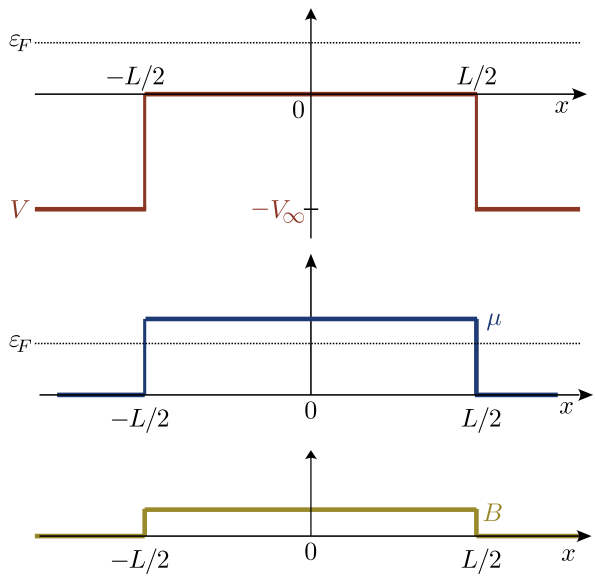}}
\caption{\label{fig_potential}
Profile of the potential $V$ (upper panel), mass $\mu$ (middle panel), and magnetic field $B$ (lower panel) along the graphene sheet. A nonzero mass is included for the sake of generality, but the case $\mu=0$ is our main interest. The contact regions $|x|>L/2$ are modeled by a deep potential well (depth $V_{\infty}\gg \hbar v/L$). Spin precession in the contacts is neglected, so we set $B=0$ there. For charge neutrality in the region $|x|<L/2$ the Fermi energy is lowered to $\varepsilon_{F}=0$ (Dirac point). 
}
\end{figure}

Following Ref.\ \cite{Two06}, the contact regions are modeled by a deep potential well, $V=-V_{\infty}\theta(|x|-L/2)$. (We will eventually take the limit $V_{\infty}\rightarrow\infty$.) The Fermi energy is tuned to the Dirac point $\varepsilon_{F}=0$ in the region between the contacts $|x|<L/2$. There are therefore no propagating modes in this region, while the contacts support a large number $N_{\infty}=k_{\infty}W/\pi$ of propagating modes (with $k_{\infty}=V_{\infty}/\hbar v$ the Fermi wave number in the contact region).

Our main interest is in the case of massless electrons, but since carriers in graphene may acquire a mass for certain substrates \cite{Gio07,Zho07}, we will include a possible nonzero mass term in the calculations. The effect of a mass is only important near the Dirac point, so we may set the mass to zero in the contact regions, taking the mass profile $\mu\theta(L/2-|x|)$ shown in Fig.\ \ref{fig_potential}.

The electron spin precesses in the $y-z$ plane around the magnetic field $\bm{B}=B\hat{x}$. We assume that the length $L$ of the region between the contacts is large compared to the length of the contacts themselves, so that we may neglect the precession in the contact region and take the magnetic field profile $B\theta(L/2-|x|)$. 

The Hamiltonian is given by
\begin{equation}\label{genHam}
 H=\mathbb{I}_s\otimes H_D -\tfrac{1}{2}\hbar\omega_{B} \sigma_x \otimes \mathbb{I}_{ps},
\end{equation}
where $\mathbb{I}_s$ and $\mathbb{I}_{ps}$ are identity matrices in real spin space and in pseudospin space, respectively. The Pauli matrix $\sigma_{x}$ in the second term acts on the real spin and accounts for the Zeeman energy, with $\omega_{B}=g\mu_{B}B$ the Larmor frequency, $\mu_{B}$ the Bohr magneton, and $g\approx 2$ the gyromagnetic factor. The first term contains the Dirac Hamiltonian, 
\begin{equation}\label{DiracHam}
 H_D=v(\sigma_x p_x+\sigma_y p_y)+\sigma_z \mu+V,
\end{equation}
for a single valley in graphene (no intervalley scattering). The Pauli matrices $\sigma_{x},\sigma_{y},\sigma_{z}$ in $H_D$ act on the pseudospin (or sublattice) degree of freedom. We neglect the coupling between the real spin and the orbit, which is weak in graphene.

We seek the currents $I_{31}$ and $I_{41}$ flowing from contact $1$ into contacts $3$ and $4$ separated by a distance $L$. These are determined by the transmittances ${\cal T}_{\downarrow\uparrow}$ and ${\cal T}_{\uparrow\uparrow}$ with and without spin flip:
\begin{equation}
I_{31}={\cal T}_{\downarrow\uparrow}G_{0}V_{1},\;\;I_{41}={\cal T}_{\uparrow\uparrow}G_{0}V_{1}.\label{IcalT}
\end{equation}
(The conductance quantum $G_{0}=2e^{2}/h$ accounts for a two-fold valley degeneracy.)

For any precessing spin, the probability of a spin-flip after a time $t$ is $\frac{1}{4}(\omega_{B}t)^{2}$ to second order in $B$. This suggests the definition of an effective transmission time $\tau^{(2)}$, in terms of the fraction ${\cal T}_{\downarrow\uparrow}/({\cal T}_{\uparrow\uparrow}+{\cal T}_{\downarrow\uparrow})={\cal T}_{\downarrow\uparrow}/{\cal T}_{\uparrow\uparrow}+{\cal O}(B^{4})$ of transmitted electrons that have flipped their spin:
\begin{equation}
{\cal T}_{\downarrow\uparrow}/{\cal T}_{\uparrow\uparrow}=\tfrac{1}{4}(\omega_{B}\tau^{(2)})^{2}+{\cal O}(B^{4}).\label{taudef}
\end{equation}
Our goal is to calculate this time $\tau^{(2)}$.

\section{Calculation of the transmission time from spin precession}
\label{transmissiontime}

The eigenvectors of the Hamiltonian~\eqref{genHam} corresponding to the eigenvalue $\varepsilon_F$ read
\begin{align}
& \Psi^{+}=\frac{1}{2}
\begin{pmatrix}
 1 \\
1
\end{pmatrix}\otimes
\begin{pmatrix}
 1\\
z_{k^{+}}
\end{pmatrix}
e^{i k^{+} x + i q y}\label{eigen+},\\
& \Psi^{-}=\frac{1}{2}
\begin{pmatrix}
 1 \\
-1
\end{pmatrix}\otimes
\begin{pmatrix}
 1\\
\widetilde{z}_{k^{-}}
\end{pmatrix}
e^{i k^{-} x + i q y}\label{eigen-},\\
& z_{k}=\frac{k+i q}{\varepsilon_F-V+\mu+{\cal B}},\;\; \widetilde{z}_{k}=\frac{k+i q}{\varepsilon_F-V+\mu-{\cal B}}.
\end{align}
We abbreviate ${\cal B}=\hbar\omega_{B}/2$ and set $\hbar v$ to unity (restoring units in the final expressions). The wave vectors
\begin{equation}
k^{\pm}=\sqrt{(\varepsilon_F-V\pm{\cal B})^2-\mu^2-q^2}
\end{equation}
are the longitudinal wave vectors. The wave vector $q$ is the transverse wave vector.

The left spinor in the tensor product in Eqs.~\eqref{eigen+} and~\eqref{eigen-} represents the state of the real spin and the right spinor represents the state of the pseudospin. The superscripts $+$ and $-$ indicate the spin polarization along the $x$ axis: the wave functions $\Psi^{+}$ and $\Psi^{-}$ are eigenstates of $\sigma_x \otimes \mathbb{I}_{ps}$ with eigenvalues $+1$ and $-1$,  respectively.

We solve the scattering problem with potential, mass, and magnetic field profiles as shown in Fig.~\ref{fig_potential}. In the contact regions $|x|>L/2$, where $V=-V_\infty\rightarrow\infty$, we have $z_{k^{+}}\rightarrow 1$, $z_{-k^{+}}\rightarrow -1$, $\widetilde{z}_{k^{-}}\rightarrow -1$, and $\widetilde{z}_{-k^{-}}\rightarrow 1$. We consider a wave incident on the charge-neutral region $|x|<L/2$ from ferromagnetic contact 1, so with spin up along the $y$-direction. Matching modes at $x=\pm L/2$ we arrive at the following linear equations for reflection and transmission amplitudes:
\begin{widetext}
\begin{multline}\label{equation_refl}
\begin{pmatrix}
 1 \\ i
\end{pmatrix}
\otimes
\begin{pmatrix}
 1 \\ 1
\end{pmatrix}
+r_{11}
\begin{pmatrix}
 1 \\ i
\end{pmatrix}
\otimes
\begin{pmatrix}
 1 \\ -1
\end{pmatrix}
+r_{21}
\begin{pmatrix}
 1 \\ -i
\end{pmatrix}
\otimes
\begin{pmatrix}
 1 \\ -1
\end{pmatrix}=
A_1
\begin{pmatrix}
 1 \\ 1
\end{pmatrix}
\otimes
\begin{pmatrix}
 1 \\ z_{k^{+}}
\end{pmatrix}e^{-i k^{+} L/2}\\
+A_2
\begin{pmatrix}
 1 \\ 1
\end{pmatrix}
\otimes
\begin{pmatrix}
 1 \\ z_{-k^{+}}
\end{pmatrix}e^{i k^{+} L/2}
+A_3
\begin{pmatrix}
 1 \\ -1
\end{pmatrix}
\otimes
\begin{pmatrix}
 1 \\ \widetilde{z}_{k^{-}}
\end{pmatrix}e^{-i k^{-} L/2}
+A_4
\begin{pmatrix}
 1 \\ -1
\end{pmatrix}
\otimes
\begin{pmatrix}
 1 \\ \widetilde{z}_{-k^{-}}
\end{pmatrix}e^{i k^{-} L/2},
\end{multline}
\begin{multline}\label{equation_trans}
t_{31}
\begin{pmatrix}
 1 \\ -i
\end{pmatrix}
\otimes
\begin{pmatrix}
 1 \\ 1
\end{pmatrix}+
t_{41}
\begin{pmatrix}
 1 \\ i
\end{pmatrix}
\otimes
\begin{pmatrix}
 1 \\ 1
\end{pmatrix}
=A_1
\begin{pmatrix}
 1 \\ 1
\end{pmatrix}
\otimes
\begin{pmatrix}
 1 \\ z_{k^{+}}
\end{pmatrix}e^{i k^{+} L/2}
+A_2
\begin{pmatrix}
 1 \\ 1
\end{pmatrix}
\otimes
\begin{pmatrix}
 1 \\ z_{-k^{+}}
\end{pmatrix}e^{-i k^{+} L/2}\\
+A_3
\begin{pmatrix}
 1 \\ -1
\end{pmatrix}
\otimes
\begin{pmatrix}
 1 \\ \widetilde{z}_{k^{-}}
\end{pmatrix}e^{i k^{-} L/2}
+A_4
\begin{pmatrix}
 1 \\ -1
\end{pmatrix}
\otimes
\begin{pmatrix}
 1 \\ \widetilde{z}_{-k^{-}}
\end{pmatrix}e^{-i k^{-} L/2}.
\end{multline}
\end{widetext}
The amplitudes $r_{11}$, $r_{21}$, $t_{31}$, and $t_{41}$ are the reflection and transmission amplitudes from contact $1$ to contacts $1,2,3,$ and $4$. Together with the coefficients $A_{1},A_{2},A_{3},$ and $A_{4}$ we have $8$ unknowns, determined by the $8$ independent equations contained in Eqs.~\eqref{equation_refl} and~\eqref{equation_trans}.

At the Dirac point, that is when $\varepsilon_F=0$, we find
\begin{align}
&T_{\downarrow\uparrow}\equiv|t_{31}|^2
=\frac{4 {\cal B}^{2} \kappa^{2}\sinh^2{\left(L\kappa\right)}}{\left[\kappa^{2}-{\cal B}^{2}+(\kappa^{2}+{\cal B}^{2})\cosh{\left(2L\kappa\right)}\right]^2},\label{tran_up_down}\\
&T_{\uparrow\uparrow}\equiv|t_{41}|^2
=\frac{4 \kappa^{4} \cosh^2{\left(L\kappa\right)}}{\left[\kappa^{2}-{\cal B}^{2}+(\kappa^{2}+{\cal B}^{2})\cosh{\left(2L\kappa\right)}\right]^2},\label{tran_up_up}\\
&R_{\downarrow\uparrow}\equiv|r_{21}|^2
=\frac{4 {\cal B}^{2} (\kappa^{2}+{\cal B}^{2})\sinh^4{\left(L\kappa\right)}}{\left[\kappa^{2}-{\cal B}^{2}+(\kappa^{2}+{\cal B}^{2})\cosh{\left(2L\kappa\right)}\right]^2},\label{refl_up_down}\\
&R_{\uparrow\uparrow}\equiv|r_{11}|^2
=\frac{\kappa^{2}(\kappa^{2}+{\cal B}^{2})\sinh^2{\left(2L\kappa\right)}}{\left[\kappa^{2}-{\cal B}^{2}+(\kappa^{2}+{\cal B}^{2})\cosh{\left(2L\kappa\right)}\right]^2}.\label{refl_up_up}
\end{align}
We have abbreviated $\kappa=\sqrt{q^{2}+\mu^{2}-{\cal B}^{2}}$. One can verify that $R_{\uparrow\uparrow}+R_{\downarrow\uparrow}+T_{\downarrow\uparrow}+T_{\uparrow\uparrow}=1$, as it should be. For ${\cal B}=0$ (no precession) we recover the transmission and reflection probabilities of Refs.\ \cite{Two06,Kat06}.

We apply periodic boundary conditions at $y=0$ and $y=W$. (Since we assume $W\gg L$, the choice of boundary condition does not matter for our results.) The transverse wave vector is then discretized as $q_n=2\pi n/W$, where $n=0,\pm 1,\pm 2,\ldots \frac{1}{2}N_{\infty}$ numbers the transverse modes. The transmittances $\mathcal{T}_{\downarrow\uparrow}$ and $\mathcal{T}_{\uparrow\uparrow}$ (with and without spin flip) are defined by the sum over modes of $T_{\downarrow\uparrow}$ and $T_{\uparrow\uparrow}$. For $W\gg L$ and $N_{\infty}\rightarrow\infty$ the sum over transmitted modes may be replaced by an integral over $q$: $\sum_{n}\rightarrow(W/2\pi)\int_{-\infty}^{\infty}dq$.

Expanding up to second order in ${\cal B}L=\omega_{B}L/2v$, we obtain the weak-field transmittances,
\begin{align}
\mathcal{T}_{\downarrow\uparrow}={}& {\cal B}^{2}L^2\frac{W}{\pi L}\int\limits_{0}^{\infty}du\,\frac{\tanh^2\sqrt{u^2+\xi^2}}{(u^2+\xi^2)\cosh^2\sqrt{u^2+\xi^2}},\label{2ndorder_tran_up_down}\\
\mathcal{T}_{\uparrow\uparrow}={}&\frac{W}{\pi L}\int\limits_{0}^{\infty}du\,\frac{1}{\cosh^2\sqrt{u^2+\xi^2}}\\
&+{\cal B}^2L^2\frac{W}{\pi L}\int\limits_{0}^{\infty}du\,\frac{\tanh\sqrt{u^2+\xi^2}}{\sqrt{u^2+\xi^2}\cosh^2\sqrt{u^2+\xi^2}}\\
&-{\cal B}^2L^2\frac{2W}{\pi L}\int\limits_{0}^{\infty}du\,\frac{\tanh^2\sqrt{u^2+\xi^2}}{(u^2+\xi^2)\cosh^2\sqrt{u^2+\xi^2}},\label{2ndorder_tran_up_up}
\end{align}
with $\xi=L \mu$ and $u= L q$.

\begin{figure}[tb]
\centerline{\includegraphics[width=0.8\linewidth]{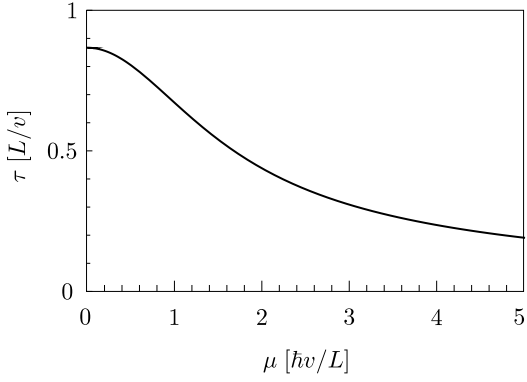}}
\caption{\label{fig_tauplot}
Dependence of the transmission time  on the mass $\mu$ of the carriers in graphene (for $W/L\gg 1$). The time is below $L/v$ for all $\mu$, as a manifestation of the Hartman effect. (While the plot is for $\tau^{(2)}$ from Eq.\ \eqref{timemass}, the time $\tau^{(1)}$ from Eq.\ \eqref{time1mass} differs only by a few percent.)
}
\end{figure}

\begin{figure}[tb]
\centerline{\includegraphics[width=0.9\linewidth]{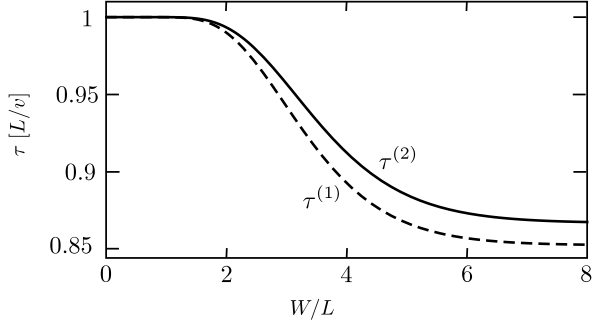}}
\caption{\label{fig_aspect}
Dependence of the transmission time  for $\mu=0$ on the aspect ratio $W/L$ of the undoped region. Both $\tau^{(1)}$ and $\tau^{(2)}$ are plotted. The limiting values for $W/L\rightarrow\infty$ are given by Eq.\ \eqref{tauresult}.
}
\end{figure}

Comparison with Eq.\ \eqref{taudef} gives an expression for the transmission time $\tau^{(2)}$, 
\begin{align}
\tau^{(2)}={}&(L/v)\left[\int\limits_{0}^{\infty}du\,\frac{\tanh^2\sqrt{u^2+\xi^{2}}}{(u^2+\xi^{2})\cosh^2\sqrt{u^2+\xi^{2}}}\right]^{1/2}\nonumber\\
&\times\left[\int\limits_{0}^{\infty}du\,\frac{1}{\cosh^2\sqrt{u^2+\xi^{2}}}\right]^{-1/2},\label{timemass}
\end{align}
plotted in Fig.\ \ref{fig_tauplot}. For massless electrons ($\xi=0$) this reduces to
\begin{equation}\label{timemassless}
\tau^{(2)}=\frac{L}{v}\left[\int\limits_{0}^{\infty}du\,\frac{\tanh^2 u}{u^2\cosh^2 u}\right]^{1/2}=0.87\,L/v,
\end{equation}
as announced in Eq.\ \eqref{tauresult}. In the large-$\mu$ limit $\tau^{(2)}\rightarrow\hbar/\mu$, independent of the distance $L$ over which the electrons are transmitted. This is the electronic analogue of the Hartman effect \cite{Har62,note1}.

These results are for aspect ratios $W/L\gg 1$, but the dependence on the aspect ratio is rather weak, as illustrated in Fig.\ \ref{fig_aspect}.

\section{The case of perpendicularly aligned magnetizations}
\label{sec_tau1}

We now turn to the case that the magnetization at the two ends of the graphene sheet is mutually perpendicular, as well as being perpendicular to the magnetic field $\bm{B}$. Referring to Fig.\ \ref{fig_setup}, we would have the magnetization in contacts $1,2$ along the $\pm y$-direction and the magnetization in contacts $3,4$ along the $\pm z$-direction (with $\bm{B}$ along $x$). The transmittances ${\cal T}_{\downarrow\uparrow}$ or ${\cal T}_{\uparrow\uparrow}$ defined in Eq.\ \eqref{IcalT} now refer to the transmission of a spin-up in the $\sigma_{y}$ basis to a spin-down or spin-up in the $\sigma_{z}$ basis.

A spin which is initially aligned along the $y$-direction, acquires after a time $t$ a polarization in the $z$-direction given by $\omega_{B}t+{\cal O}(B^{2})$. Analogously to Eq.\ \eqref{taudef}, we now define the effective transmission time $\tau^{(1)}$ by
\begin{equation}
\frac{{\cal T}_{\uparrow\uparrow}-{\cal T}_{\downarrow\uparrow}}{{\cal T}_{\uparrow\uparrow}+{\cal T}_{\downarrow\uparrow}}=\omega_{B}\tau^{(1)}+{\cal O}(B^{2}).\label{tau1def}
\end{equation}

A very similar calculation as in Sec.\ \ref{transmissiontime} gives 
\begin{align}
\tau^{(1)}={}&(L/v)\int\limits_{0}^{\infty}du\,\frac{\tanh\sqrt{u^2+\xi^{2}}}{\sqrt{u^2+\xi^{2}}\cosh^2\sqrt{u^2+\xi^{2}}}\nonumber\\
&\times\left[\int\limits_{0}^{\infty}du\,\frac{1}{\cosh^2\sqrt{u^2+\xi^{2}}}\right]^{-1}.\label{time1mass}
\end{align}
The $\mu$-dependence of $\tau^{(1)}$ is only a few percent different from that of $\tau^{(2)}$ (plotted in Fig.\ \ref{fig_tauplot}). In the limit $\xi\equiv L\mu\rightarrow 0$ of massless electrons we find
\begin{equation}\label{time1massless}
\tau^{(1)}=\frac{L}{v}\int\limits_{0}^{\infty}du\,\frac{\tanh u}{u\cosh^2 u}=\frac{7\zeta(3)}{\pi^{2}}\frac{L}{v}=0.85\,\frac{L}{v},
\end{equation}
as announced in Eq.\ \eqref{tauresult}.

\section{Comparison with Wigner-Smith delay times}\label{WignerSmith}

We wish to derive the relationship \eqref{tautau_nrelation} between the transmission time $\tau^{(p)}$ measured in spin precession and the mode-dependent Wigner-Smith delay times $\tau_{n}$. By definition, the Wigner-Smith delay times are the eigenvalues of the Wigner-Smith time-delay matrix
\begin{equation}
Q=-i\hbar S^{\dagger}\frac{dS}{d\varepsilon_F},
\end{equation}
constructed from the energy dependent scattering matrix $S$. The eigenvalues $\tau_{n}$ of $Q$ appear in certain transport properties \cite{Bro97,God99}, but they are usually not directly measurable. For example, the thermopower of a single-channel conductor depends on the difference $\tau_{1}-\tau_{2}$ of the two eigenvalues of $Q$, as well as on the eigenvectors. It is therefore not obvious \textit{a priori} that $\tau^{(p)}$ can be related to the $\tau_{n}$'s.

Since we seek the delay times in the limit of zero magnetic field, we can consider a simpler scattering problem than in the previous section, namely transmission of spinless electrons through a graphene sheet with the mass and potential profile shown in Fig.~\ref{fig_potential}. In this case the scattering matrix is given by \cite{Two06}
\begin{multline}
 S=\frac{1}{k\cos{kL}-i\varepsilon_F\sin{kL}}\\
 \times\begin{pmatrix}
   (-q-i\mu)\sin{k L} & k \\
   k & (q-i\mu)\sin{k L} \\
 \end{pmatrix},
\end{multline}
where $k=\sqrt{\varepsilon_F^2-\mu^2-q^2}$.

The general energy-dependent expression for $Q$ is lengthy, but at the Dirac point it simplifies to
\begin{equation}
Q=\tau(q)\begin{pmatrix}
1&0\\
0&1
\end{pmatrix},\;\;
\tau(q)=\frac{\tanh{(L \sqrt{q^2+\mu^2})}}{v\sqrt{q^2+\mu^2}}.\label{QDirac}
\end{equation}
So for each mode $n$ there is a single doubly degenerate Wigner-Smith delay time $\tau_{n}=\tau(q=2\pi n/W)$. The mode-dependent transmission probability at the Dirac point is $T_{n}=T(q=2\pi n/W)$ with
\begin{equation}
T(q)=\frac{1}{\cosh^{2}(L \sqrt{q^2+\mu^2)}}.\label{TDirac}
\end{equation}

Combination of Eqs.\ \eqref{QDirac} and \eqref{TDirac} shows that
\begin{align}
\frac{\sum_{n}\tau_{n}^{p}T_{n}}{\sum_{n}T_{n}}={}&\int_{0}^{\infty}dq\,\frac{\tanh^{p}{(L \sqrt{q^2+\mu^2})}}{v^{p}(q^2+\mu^2)^{p/2}\cosh^{2}(L \sqrt{q^2+\mu^2)}}\nonumber\\
&\times\left[\int_{0}^{\infty}dq\,\frac{1}{\cosh^{2}(L \sqrt{q^2+\mu^2)}}\right]^{-1},\label{tautau_nrelation2}
\end{align}
where we have replaced the sum over modes by an integration over wave vectors (appropriate for $W/L\gg 1$). Comparison with the expressions \eqref{timemass} and \eqref{time1mass} for $\tau^{(2)}$ and $\tau^{(1)}$ proves the identity \eqref{tautau_nrelation} of the transmission time measured in spin precession and the weighted average of the mode-dependent Wigner-Smith delay times. 

\section{Conclusion}
\label{discuss}

In conclusion, we have shown how spin precession in graphene may reveal an unusual dynamical aspect of ballistic quantum transport at the Dirac point. In a clean charge-neutral graphene sheet of length $L$, the transmission is via evanescent rather than propagating waves. While for propagating waves the transmission time is bounded by $\tau\geq L/v$, evanescent waves have no well-defined velocity and can show a shorter $\tau$ in a precession measurement. This is the electronic analogue of the Hartman effect from optics \cite{Har62,Win06}. Our result \eqref{tauresult} for massless electrons is not much below $\tau=L/v$, but it does provide an unambiguous demonstration of this apparent superluminality.

From a conceptual point of view, our analysis demonstrates, firstly, that the pseudodiffusive aspects of ballistic transmission at the Dirac point (as observed in conductance and shot noise \cite{Mia07,Dan08,DiC08}), are restricted to static properties. The dynamics is not diffusive in any sense (no $L^{2}$ scaling of $\tau$). Secondly, our analysis demonstrates via the relation \eqref{tautau_nrelation} that the Wigner-Smith delay times are directly observable through spin precession at the Dirac point.

We finally notice a qualitative difference between spin precession in a tunnel barrier and spin precession at the Dirac point. As pointed out by B\"{u}ttiker \cite{But83}, the spin of a tunneling electron not only precesses in the $y-z$ plane perpendicular to $\bm{B}$, but in addition aligns itself along the magnetic field. The rotation of the spin out of the $y-z$ plane (dominant in a tunnel barrier, but ignored in the Larmor clock \cite{Baz67,Ryb67}) appears because of a difference in tunnel probabilities for spins parallel or antiparallel to $\bm{B}$. No such out-of-plane rotation appears at the Dirac point, due to the fact that the energy-dependent transmission probabilities are extremal at zero energy. The spin precession geometry analyzed in this work is therefore particularly close to the original concept of a Larmor clock.

\acknowledgments
We benefited from discussions with A. R. Akhmerov, M. B\"{u}ttiker, and B. J. van Wees. This project was supported by the Dutch Science Foundation NWO/FOM and by an ERC Advanced Investigator Grant.


\begin{thebibliography}{99}
\bibitem{note1} The effect discovered by Hartman is superluminality in the $L\rightarrow\infty$ limit. Since the term ``superluminality'' is inappropriate for graphene, where all velocities remain well below the speed of light, we avoid this term and speak of the Hartman effect also for finite $L$.
\bibitem{note2} An alternative way to measure the precession time, closer to the geometry used in Refs.\ \cite{Tom07,Pop09}, is to ground only contacts $2$ and $4$, while contact $3$ is a voltage probe (drawing no current). The ratio ${\cal T}_{\downarrow\uparrow}/{\cal T}_{\uparrow\uparrow}$ that we need to obtain $\tau^{(2)}$ can then be measured as the ratio $V_{3}/V_{1}$.
\end{thebibliography}
\end{document}